\documentclass[a4paper,11t]{article}
\usepackage[top=2.54cm,bottom=2.54cm, left=2cm, right=2cm]{geometry}

\usepackage{amsmath}
\usepackage{amsfonts}
\usepackage{amssymb}
\usepackage{graphicx}
\usepackage{bm}
\usepackage{enumerate}
\usepackage{color}
\usepackage[sort,comma]{natbib}
\usepackage{float}
\usepackage[colorlinks=true, allcolors=blue]{hyperref}
\usepackage{changepage}
\usepackage{dirtytalk}
\usepackage{pdflscape}
\usepackage{booktabs}
\usepackage[normalem]{ulem}
\newcommand{\dotsim}{\mathrel{\dot{\sim}}}

\title{Directional data analysis using the spherical Cauchy and the Poisson kernel-based distribution}
\author{Michail Tsagris$^1$, Panagiotis Papastamoulis$^2$ and Shogo Kato$^3$ \\
\\
$^1$ Department of Economics, University of Crete, Gallos Campus, Rethymnon, Greece \\ 
\href{mailto:mtsagris@uoc.gr}{mtsagris@uoc.gr} \\
$^2$ Department of Statistics, Athens University of Economics and Business,
Athens, Greece \\
\href{mailto:papastamoulis@aueb.gr}{papastamoulis@aueb.gr} \\
$^2$ Institute of Statistical Mathematics, Tokyo, Japan \\
\href{mailto:skato@ism.ac.jp}{skato@ism.ac.jp}
}

\begin{document}

\maketitle

\begin{center}
\textbf{Abstract}
\end{center}
In 2020, two novel distributions for the analysis of directional data were introduced: the spherical Cauchy distribution and the Poisson kernel-based distribution. This paper provides a detailed exploration of both distributions within various analytical frameworks. To enhance the practical utility of these distributions, alternative parametrizations that offer advantages in numerical stability and parameter estimation are presented, such as implementation of the Newton-Raphson algorithm for parameter estimation, while facilitating a more efficient and simplified approach in the regression framework. Additionally, a two-sample location test based on the log-likelihood ratio test is introduced. This test is designed to assess whether the location parameters of two populations can be assumed equal. The maximum likelihood discriminant analysis framework is developed for classification purposes, and finally, the problem of clustering directional data is addressed, by fitting finite mixtures of Spherical Cauchy or Poisson kernel-based distributions. Empirical validation is conducted through comprehensive simulation studies and real data applications, wherein the performance of the spherical Cauchy and Poisson kernel-based distributions is systematically compared. 

\section{Introduction}
Directional data refers to multivariate data constrained to a unit norm, with the sample space represented as:
\begin{eqnarray*}
\mathbb{S}^d = \left\lbrace \mathbf{x} \in \mathbb{R}^{d+1} \bigg\vert \left|\left|\mathbf{x}\right|\right| = 1 \right\rbrace,
\end{eqnarray*}
where $\left|\left| \cdot \right|\right|$ denotes the Euclidean norm. For \(d=1\), circular data reside on a circle, while for \(d=2\), spherical data are situated on a sphere. 

Circular data are prevalent in various fields, including political science \citep{gill2010}, criminology \citep{shirota2017}, biology \citep{landler2018}, ecology \citep{horne2007}, and astronomy \citep{soler2019}. Conversely, spherical data are encountered in disciplines such as geology \citep{chang1986}, environmental sciences \citep{heaton2014}, image analysis \citep{straub2015}, robotics \citep{bullock2014}, and space exploration \citep{kent2016}, to name a few.

Historically, numerous spherical and hyperspherical distributions have been developed, with the von Mises-Fisher distribution \citep{fisher1953} and the projected normal distribution \citep{kendall1974} among the earliest and most commonly used. More recent contributions include the spherical Cauchy (SC) distribution \citep{kato2020} and the Poisson kernel-based (PKB) distribution \citep{golzy2020}. Although these distributions assume rotational symmetry, which may limit their applicability in certain contexts, they have demonstrated effectiveness in various scenarios, particularly for spherical data \citep{tsagris2019, tsagris2024a}.

This paper investigates the SC \citep{kato2020} and PKB \citep{golzy2020} distributions across five key aspects: random vector simulation, maximum likelihood estimation, hypothesis testing about the location parameters, regression modeling, and discriminant analysis. While existing distributions address these areas, they often come with limitations, such as computational inefficiency or inadequate regression models.

In terms of random vector simulation, many distributions, such as the von Mises-Fisher \citep{wood1994} and Kent \citep{kent2018} distributions, rely on rejection sampling. Other distributions, including the projected normal, the elliptically symmetric angular Gaussian (ESAG) \citep{paine2018}, and the spherical projected Cauchy \citep{tsagris2024b}, avoid this approach. The SC distribution is shown to be straightforward to simulate, whereas the PKB distribution requires rejection sampling, which may be a drawback when rapid random vector generation is necessary.

Efficient maximum likelihood estimation (MLE) is crucial for both simulation studies and the analysis of large-scale data. The mean direction of the von Mises-Fisher distribution has a closed-form estimate, and its concentration parameter can be estimated using a fast Newton-Raphson (NR) algorithm \citep{sra2012}. In contrast, parameter estimation for the Kent and projected normal distributions is more complex and typically relies on slower numerical optimizers, especially in higher dimensions. The SC and PKB distributions benefit from either a NR-based MLE or a hybrid algorithm. An additional perk of both distributions is that the normalizing constant is available in closed form.

Various hypothesis testing procedures for comparing mean directions have been proposed. \cite{tsagris2024a} compared methods for two populations, using both asymptotic theory and computational techniques to obtain p-values. Tests assuming homogeneity among samples often failed to maintain the nominal type I error rate, while tests not reliant on this assumption performed more accurately. The proposed log-likelihood ratio tests for the SC and PKB distributions do not assume equal concentration parameters, addressing this issue.

\cite{presnell1998} examined the relationship between the mean direction of the projected normal distribution on the circle and covariates, allowing the concentration parameter to vary among residuals, analogous to heteroscedasticity in linear regression. This concept was later applied by \cite{paine2020} to the ESAG, Kent, and von Mises-Fisher distributions, and by \cite{tsagris2024b} to the projected Cauchy distribution. This strategy is also applied here to the SC and PKB distributions.

Discriminant analysis, or supervised learning, is another significant topic. \cite{tsagris2019} compared maximum likelihood-based classifiers with the k-NN algorithm, demonstrating that rotationally symmetric distributions can perform as well as or better than elliptically symmetric distributions. The SC and PKB distributions are found to perform comparably in discriminant analysis, with the SC exhibiting faster performance.

The final topic of research is model based clustering. \cite{hornik2014} proposed mixtures of von Mises-Fisher distributions to this end. We address this issue by estimating finite mixtures of SC and PKB distributions, under a maximum likelihood framework. For this purpose we implement the Expectation-Maximization algorithm \citep{dempster1977}.  

The rest of the paper is organized as follows. Section 2 presents the framework for random generation (Section \ref{sec:sim}), maximum likelihood estimation (Section \ref{mle}), hypothesis testing (Section \ref{sec:ht}), regression analysis (Section \ref{sec:reg}), classification through maximum likelihood discriminant analysis (Section \ref{sec:clas}) and model-based clustering (Section \ref{sec:mix}). Section \ref{simulations} evaluates the performance of these distributions through simulations, while Section \ref{real} demonstrates their effectiveness with real data. Finally, Section \ref{conclusions} concludes the paper.

\section{The spherical Cauchy and Poisson kernel-based distributions} \label{sc}
A model that appears to be closely related to the classical vMF distribution is the SC distribution \citep{kato2020}, which can be seen as the generalisation of the wrapped Cauchy distribution \citep[pg.~50--52]{mardia2000} to the sphere (and hyper-sphere). 

The density of the SC on $\mathbb{S}^d$ is given by
\begin{eqnarray} \label{sc1}
f(\bm{y})=C_d\left(\frac{1-\rho^2}{1+\rho^2-2\rho\bm{y}^\top{\bf m}}\right)^d,  
\end{eqnarray}
where ${\bf m} \in \mathbb{S}^d$ is the location direction, that controls the mode of the density, $\rho \in [0, 1)$ plays the role of the concentration parameter and $C_d=\frac{\Gamma\left[(d+1)/2\right]}{2\pi^{(d+1)/2}}$ is the normalizing constant. For the most part of this paper though we will use an alternative parameterization that was used in \cite{tsagris2024b} 
\begin{eqnarray} \label{sc2}
f(\bm{y}) =C_d\left(\sqrt{\|\pmb{\mu}\|^2+1}-\bm{y}^\top\pmb{\mu}\right)^{-d}=
C_d\left(\sqrt{\gamma^2+1}-\alpha\right)^{-d},
\end{eqnarray}
where $\bm{\mu} \in \mathbb{R}^{d+1}$ is the unconstrained location parameter, $\alpha=\bm{y}^\top\pmb{\mu}$, $\gamma=\|\pmb{\mu}\|$, ${\bf m}=\pmb{\mu}/\gamma$ and $\rho=(\sqrt{\gamma^2+1}-1)/\gamma$ ($\gamma=\frac{2\rho}{1-\rho^2}$). The benefit of this parameterization is that the maximisation with respect to the location parameter is unconstrained. 

A second, similar, distribution is the PKB distribution that was proposed by \cite{golzy2020}, and can also be seen as the the generalisation of the wrapped Cauchy distribution, whose density is given by
\begin{eqnarray} \label{pkbd1}
f(\bm{y})=C_d\frac{1-\rho^2}{\left(1+\rho^2-2\rho\bm{y}^\top{\bf m}\right)^{(d+1)/2}},
\end{eqnarray}

One may express this distribution in an alternative way, similar to the SC, as a function of the unconstrained location parameter $\pmb{\mu}$  
\begin{eqnarray} \label{pkbd2}
f(\bm{y})=C_d\left(\sqrt{\gamma^2+1}-\alpha\right)^{-(d+1)/2}\left(1-\frac{(\sqrt{\gamma^2+1}-1)^2}{\gamma^2}\right)^{-(d-1)/2}.  
\end{eqnarray}

\begin{figure}[!ht]
\centering
\begin{tabular}{cc}
\includegraphics[scale = 0.55]{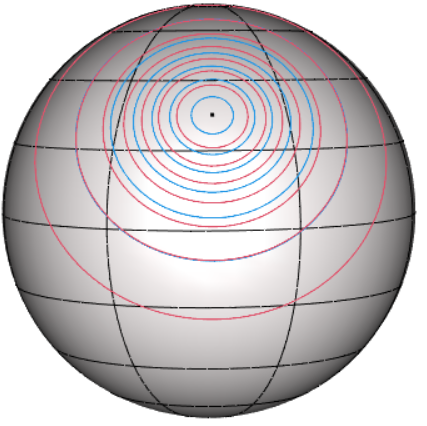}  &
\includegraphics[scale = 0.55]{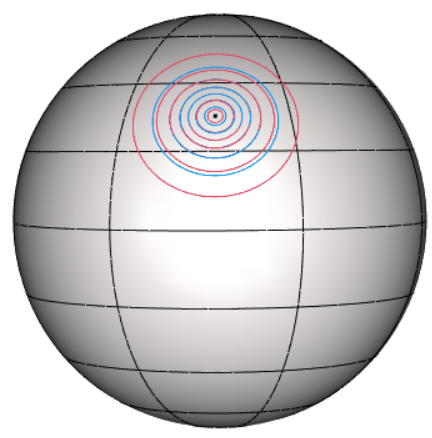}  \\
(a) $\rho=0.6$  &  (b) $\rho=0.8$
\end{tabular}
\caption{Contour plots of the SC (blue lines) and PKB (red lines) distributions with $\pmb{\mu}=(5.843,3.057,3.758)^\top$ or ${\bf m}=\left(0.770, 0.403, 0.495\right)^\top$ and two different values of $\rho$.}
\label{contour}
\end{figure}

Figure \ref{contour} presents the contour plots for some location parameter and two values of the concentration parameter $\rho$ for the SC distribution. The contour plots have the same shape, but the tails of the PKB decay slower than those of the SC distribution. The density value of the SC will be larger than the density value of the PKB, that is, if $\bm{y}^\top\pmb{\mu}<(1 + \rho^2)/2$. 

\subsection{Simulation of directional vectors}\label{sec:sim}
In order to simulate random directional vectors $\bm{y}_i$ from the $\text{SC}(\bm{y};{\bf m},\rho)$, \cite{kato2020} proposed a simple procedure that relies upon the uniform distribution and then perform some simple calculations. The fact that no rejection sampling is necessary, a strategy common with other distributions, is an appealing one. The two steps of the algorithm are described below.
\begin{enumerate}
\item Generate vectors ${\bf u}_i$ ($i=\ldots,n$) from the uniform distribution in $\mathbb{S}^d$.
\item Set $\bm{y}_i=\frac{\left({\bf u}_i + \rho {\bf m}\right)\left(1-\rho^2\sum_{j=1}^{d+1}{\bf m}_j^2\right)}{\sum_{j=1}^{d+1}\left({\bf u}_i + \rho {\bf m}_j \right)^2} + \rho {\bf m}$.  
\end{enumerate}

In order to simulate random directional vectors $\bm{y}_i$ from the $\text{PKB}(\bm{y};{\bf m},\rho)$, \cite{sablica2023} proposed a rejection sampling that is computationally more expensive than the simulation of the SC distribution. 
\begin{enumerate}
\item[Step 1] Set $\lambda = 2 \rho / (1 + \rho^2)$.
\item[Step 2] Define  $\omega_d(\lambda, \beta) = 0.5 (d+1) \log{\frac{1 + \sqrt{1 - \lambda^2}}{1 + \sqrt{1 - \lambda^2 / \beta}} } - 0.5 log{(1-\beta)}$ and find the $\beta^*$ that minimizes $\omega_d(\lambda, \beta)$ in the interval $(\lambda(2\lambda-1), 1)$.
\item[Step 4] Set $\beta_1 = \beta^* / (1 - \beta^*)$ and $\beta_2 =  - 1 + 1 / \sqrt{(1 - \beta^*)}$
\item[Step 5] Simulate $u \sim U(0,1)$ and ${\bf z}=\left(z_1,\ldots,z_{d+1}\right)$, where each $z_i$ is simulated from a standard normal distribution, $z_i \sim N(0,1)$.
\item[Step 6] Set $q=\left(\pmb{\mu}^\top{\bf z} + \beta_2\pmb{\mu}^\top{\bf z} \right)/
\sqrt{{\bf z}^\top{\bf z}+\beta_1(\pmb{\mu}^\top{\bf z})^2}$.
\item[Step 7] If $log{(u)} \leq \frac{d+1}{2}\left[ - \log{(1 - \lambda * q)} + \log{(1 - \beta^* q^2)}-log{\frac{2}{1 + \sqrt{1 - \lambda^2/\beta^*}}} \right]$ \\ 
set ${\bf x} \leftarrow \left({\bf z} + \beta_2\pmb{\mu}^\top{\bf z}\pmb{\mu} \right)/
\sqrt{{\bf z}^\top{\bf z}+\beta_1(\pmb{\mu}^\top{\bf z})^2}$, otherwise return to Step 5.
\end{enumerate}

\subsection{Maximum likelihood estimation} \label{mle}
The log-likelihood for a sample of directional vectors $\bm{y}_i$, $i=1\ldots,n$ of the SC distribution, using Eq. (\ref{sc2}) is given by 
\begin{eqnarray} \label{lik2}
\ell_{SC} = n\log{C_d} - d\sum_{i=1}^n\log\left(\sqrt{\gamma^2+1}-\alpha_i\right),
\end{eqnarray}
where $C_d$ denotes the normalizing constant. To perform maximum likelihood estimation (MLE) the Newton-Raphson (NR) algorithm can be employed, to maximise $\ell_{SC}$ (\ref{lik2}). The starting value for the NR algorithm is the sample mean vector for which the log-likelihood value is computed. At each successive step the estimated mean vector is updated via $\pmb{\mu}^{t+1}=\pmb{\mu}^t - {\bf H}^{-1}{\bf J}$ and the algorithm terminates when the difference between two successive log-likelihood values is smaller than some tolerance value $\epsilon$ (for instance $\epsilon=10^{-6}$). 

In the NR algorithm, the concentration parameter $\rho$ is embedded within the estimation of the mean vector $\pmb{\mu}$. An alternative optimization strategy that disentangles the mean direction $\bf m$ from the concentration parameter, that is in similar spirit with the maximization step adopted by \cite{golzy2020}, is also proposed. The method is a hybrid of the Brent algorithm and of the fixed points iteration algorithm. 

The relevant log-likelihood (excluding the normalizing constant) of the parameterization in Eq. (\ref{sc1}) is given by
\begin{eqnarray} \label{lik1}
\ell_{SC} = n\log{C_d} + nd\log\left(1 - \rho^2\right) - d\sum_{i=1}^n\log\left(1+ \rho^2 - 2\rho \bm{y}_i^\top {\bf m} \right).
\end{eqnarray}
The steps of the hybrid algorithm are delineated below.
\begin{enumerate}
\item[Step 1] Start with an initial mean direction given by $\hat{\bf m}=\frac{\bar{\bm{y}}}{\|\bar{\bm{y}}\|}$, where $\bar{\bm{y}}$ denotes the sample mean vector.
\item[Step 2] Using this mean direction obtain $\hat{\rho}$ that maximises the log-likelihood in Eq. (\ref{lik1}), using the Brent algorithm \citep{brent1973}.
\item[Step 3] For the estimated $\hat{\rho}$ from Step 2, update the mean direction, using the fixed points iteration algorithm, by maximising the log-likelihood in Eq. (\ref{lik1}) under the constraint that the mean direction lies in $\mathbb{S}^{d-1}$. The Lagrangian function takes the following form
\begin{eqnarray} \label{lagrange}
\ell_{SC} = n\log{C_d} + nd\log\left(1 - \hat{\rho}^2\right) - d\sum_{i=1}^n\log\left(1+ \hat{\rho}^2 - 2\hat{\rho} \bm{y}_i^\top {\bf m} \right) + \lambda \left({\bf m}^\top{\bf m} - 1 \right).
\end{eqnarray}
Equating the derivative of (\ref{lagrange}), with respect to $\bf m$, to zero, yields
\begin{eqnarray*}
\frac{\partial \ell_{SC}}{\partial {\bf m}}= d\sum_{i=1}^n\frac{2\rho \bm{y}_i}{1+ \hat{\rho}^2 - 2\hat{\rho} \bm{y}_i^\top {\bf m}} +2\lambda{\bf m} = {\bf 0}.  
\end{eqnarray*}
The updated mean direction is given by the unit vector parallel to $\sum_{i=1}^n\frac{\hat{\rho} \bm{y}_i}{1+ \hat{\rho}^2 - 2\hat{\rho} \bm{y}_i^\top \hat{{\bf m}}}$.
\item[Step 4] Repeat Steps 2-3 until the log-likelihood in Eq. (\ref{lik1}) improves no more than some tolerance value $\epsilon$.
\end{enumerate}

The strategy employed in Step 3 was also employed by \cite{cabrera1990} in order to estimate the median direction and by \cite{fayomi2024} to obtain the eigenvectors of the Cauchy principal component analysis.  

As for the PKB distribution, the second representation (\ref{pkbd2}) eases the calculations, since the log-likelihood can be written as
\begin{eqnarray} \label{pkblik2}
\ell_{PKB} = n\log{\left(C_d\right)} -\frac{d+1}{2}\sum_{i=1}^n \log{\left(\sqrt{\gamma^2+1}-\alpha_i\right)} - n\frac{d-1}{2}\left[\log{2} + \log{\left(\sqrt{\gamma^2+1}-1\right)} - \log{\gamma^2}\right]. 
\end{eqnarray}
The derivatives are very similar to those of the SC log-likelihood, with the exception of some extra terms. As for the hybrid algorithm, the mathematics are nearly the same in the SC case. Lastly, we highlight that an advantage of the hybrid method over the NR method is that the first may also be used in the $n<d$ case. 

\subsection{Log-likelihood ratio test for equality of two location parameters}\label{sec:ht}
In order to test the equality of equal population location parameters, based on two samples\footnote{Evidently, the procedure can be generalised to the case of more than two groups.} that follow the SC distribution we will employ the log-likelihood ratio test, without assuming equality of the concentration parameters. Under the null hypothesis one must maximise the following log-likelihood (ignoring $C_d$) with respect to the common mean direction ${\bf m}_c$ and the two concentration parameters $\rho_1$ and $\rho_2$. 
\begin{eqnarray} \label{likh0}
\ell_0  = n_1d\log\left(1 - \rho_1^2\right) - d\sum_{i=1}^{n_1}\log\left(1+ \rho_1^2 - 2\rho_1 \bm{y}_{1i}^\top {\bf m}_c \right) + n_2d\log\left(1 - \rho_2^2\right) - d\sum_{i=1}^{n_2}\log\left(1+ \rho_2^2 - 2\rho_2 \bm{y}_{2i}^\top {\bf m}_c \right),        
\end{eqnarray}
where $n_j$ and $\bm{y}_{ji}$ refer to the $j$-th sample size and the $i$-th observation of sample $j$, respectively, for $j=1,2$. Under the alternative hypothesis, the two location parameters are not equal and hence the relevant log-likelihood to be maximised is given by
\begin{eqnarray} \label{likh1}
\ell_1  = n_1d\log\left(1 - \rho_1^2\right) - d\sum_{i=1}^{n_1}\log\left(1+ \rho_1^2 - 2\rho_1 \bm{y}_{1i}^\top {\bf m}_1 \right) + n_2d\log\left(1 - \rho_2^2\right) - d\sum_{i=1}^{n_2}\log\left(1+ \rho_2^2 - 2\rho_2 \bm{y}_{2i}^\top {\bf m}_2 \right),        
\end{eqnarray}
where ${\bf m}_j$ refers to the location parameter of the $j$-th sample. In order to maximise $\ell_0$ (\ref{likh0}) the hybrid maximisation approach is used, whereas for the maximisation of $\ell_1$ (\ref{likh1}) this is accomplished via the NR algorithm, applied to each sample separately. If the null hypothesis is true, standard likelihood theory states that $\Lambda = 2\left[\ell_1(\hat{\bf m}_1, \hat{\bf m}_2, \hat{\rho}_1,\hat{\rho}_2)-\ell_0(\tilde{\bf m}_c, \tilde{\rho}_1,\tilde{\rho}_2)\right] \dotsim \chi^2_d$, where $\hat{\bf m}_1, \hat{\bf m}_2, \hat{\rho}_1,\hat{\rho}_2$ denote the estimated parameters under $H_1$, while $\tilde{\bf m}_c, \tilde{\rho}_1,\tilde{\rho}_2$ denote the estimated parameters under $H_0$.

The same strategy was adopted for the case of the PKB distribution as well, with the exception that the formulas are slightly different.

\subsection{Regression analysis}\label{sec:reg}
When a set of $p$ covariates is present, we link the location parameter to the covariates by $\bm{\mu}_i= {\bf B}^\top {\bm x}_i$, where ${\bf B}=\left(\bm{\beta}_1,\ldots,\bm{\beta}_{d+1} \right)$ denotes the matrix of the regression coefficients, $\bm{\beta}_i \in \mathbb{R}^p$, and ${\bm x}_i \in \mathbb{R}^p$ denotes a column of the design matrix. The relevant log-likelihood, using (\ref{sc2}), becomes 
\begin{eqnarray} \label{screg}
\ell_{SC} = n \log\left(C_d\right) - d\sum_{i=1}^n\log\left(\sqrt{\gamma_i^2+1}-\alpha_i\right), 
\end{eqnarray}
where $\alpha_i= \bm{y}_i^\top\bm{\mu}_i$ and $\gamma_i=\|\bm{\mu}_i\|$. 

The advantage of the parametrization in Eq. (\ref{sc2}) is evident in the regression setting. Similarly to \cite{paine2020} and \cite{tsagris2024b} the errors are heteroscedastic, because we do not assume a common concentration parameter. Each directional vector has its own concentration parameter $\gamma_i$ that is linked to $\rho_i$ as mentioned earlier. The matrix of the regression coefficients is again estimated via the NR algorithm. 

The asymptotic distribution of the maximum likelihood estimator for this regression model is available.
Let $\bm{\beta} = (\bm{\beta}_1^\top , \ldots, \bm{\beta}_{d+1}^\top )^\top$ denote the $(d+1)p$-dimensional parameter vector obtained by vectorizing the matrix $\bf{B}$.
Then, under certain regularity conditions, the following holds for the maximum likelihood estimator $\hat{\bm{\beta}}_{SC}$ of $\bm{\beta}$:
\begin{equation}
\sqrt{n} \, {\cal I}_{SC}^{1/2} \left( \hat{\bm{\beta}}_{SC} - \bm{\beta} \right) \stackrel{{\rm d}}{\longrightarrow} N \left( \bm{0} , {\bf I}_{(d+1)p} \right) \quad (n \rightarrow \infty). \label{eq:asy_normal_sc}
\end{equation}
Here ${\cal I}_{SC}$ is the observed Fisher information matrix given by
\begin{align}
\begin{split} \label{eq:fisher_sc}
{\cal I}_{SC} = & \ \frac{d}{n} \sum_{i=1}^n \Biggl[ \frac{ {\bf I}_{d+1} \otimes \bm{x}_i \otimes \bm{x}_i^\top - \sqrt{\gamma_i^2+1} \left( \bm{\mu}_i \otimes \bm{x}_i \right) \left(  \bm{\mu}_i \otimes \bm{x}_i \right)^\top }{ \sqrt{\gamma_i^2+1} \left( \sqrt{\gamma_i^2+1} - \alpha_i \right)} \\
 & -  \frac{ \left\{ \left( \bm{\mu}_i - \sqrt{\gamma_i^2+1} \, \bm{y}_i \right) \otimes \bm{x}_i \right\} \left\{ \left( \bm{\mu}_i - \sqrt{\gamma_i^2+1} \, \bm{y}_i \right) \otimes \bm{x}_i \right\}^\top  }{ (\gamma_i^2+1) \left( \sqrt{\gamma_i^2+1} - \alpha_i \right)^2} \Biggr],
 \end{split}
\end{align}
where $\otimes$ denotes the Kronecker product.

The log-likelihood of the PKB regression model, using the parameterization from Eq. (\ref{pkbd2}) is written as follows
\begin{eqnarray} \label{pkbdreg}
\ell_{PKB}=n\log{\left(C_d\right)} - n\frac{d-1}{2}\log{2} -\frac{d+1}{2}\sum_{i=1}^n \log{\left(\sqrt{\gamma_i^2+1}-\alpha_i\right)} -\frac{d-1}{2}\left[\sum_{i=1}^n\log{\left(\sqrt{\gamma_i^2+1}-1\right)} - \sum_{i=1}^n\log{\gamma_i^2}\right],  
\end{eqnarray}
where $\alpha_i$ and $\gamma_i$ are the same as in the case of the SC regression.
Asymptotic normality also holds for the maximum likelihood estimator of the PKB regression model under regularity conditions.
Let $\hat{\bm{\beta}}_{PKB}$ denote the maximum likelihood estimator of $\bm{\beta}$ for the PKB regression. Then, we have
\begin{equation}
\sqrt{n} \, {\cal I}_{PKB}^{1/2} \left( \hat{\bm{\beta}}_{PKB} - \bm{\beta} \right) \stackrel{{\rm d}}{\longrightarrow} N \left( \bm{0} , {\bf I}_{(d+1)p} \right) \quad (n \rightarrow \infty), \label{eq:asy_normal_pkb}
\end{equation}
where the observed Fisher information matrix ${\cal I}_{PKB}$ is
\begin{align}
\begin{split} \label{eq:fisher_pkb}
 {\cal I}_{PKB} = \ & \frac{d+1}{2d} {\cal I}_{SC} + \frac{d-1}{2n} \sum_{i=1}^n \Biggl[ \frac{\sqrt{\gamma_i^2+1} -1 }{\gamma_i^2} {\bf I}_{d+1} \otimes \bm{x}_i^\top \otimes \bm{x}_i \\
  & + \frac{3 \gamma_i^2 - 2 \sqrt{\gamma_i^2+1} +2}{\gamma_i^4 (\gamma_i^2+1)} \left( \bm{\mu}_i \otimes \bm{x}_i \right) \left( \bm{\mu}_i \otimes \bm{x}_i \right)^\top \Biggr].
  \end{split}
\end{align}

\subsection{Discriminant analysis}\label{sec:clas}
Assume there are $J$ groups in the population of interest and that given the class variable $C_i\in\{1,\ldots,J\}$, $\bm{Y}_i$ is distributed as 
\[
\bm{Y}_i|(C_i = j) \sim f(\cdot; \bm{\mu}_j)
\]
independent for $i = 1,\ldots,n$, where $\pmb{\mu}_j\in\mathbb R^{d+1}$ denotes the  (unconstrained)  parameter of the $j$-th group,  $j=1,\ldots,J$. Finally, $f(\cdot|\bm{\mu})$ denotes the probability density function (which can be either the SC or the PKB distribution) with parameter $\bm{\mu}\in\mathbb R^{d+1}$.

Furthermore, we assume that we have at hand a labeled directional dataset, that is, for each observation $\textbf{y}_i\in \mathbb{S}^d$, the allocation variable $c_i\in\{1,\ldots,J\}$ is also recorded, $i=1,\ldots,n$. The goal is to assign a feature observation $\textbf{y}_0$ into one among the $J$ classes. Under the maximum likelihood discriminant analysis framework \citep[301--305]{mardia1979}, the rule is to allocate a new observation vector $\bm{y}_0 \in \mathbb{S}^d$ in the group whose likelihood is maximized, i.e,
\[c_0 = \mathrm{argmax}_j \{f(\bm{y}_0;\bm{\mu}_j);j=1\ldots,J\}.\]
Hence, in the case of the SC, compute the following quantity  
\begin{eqnarray*}
\log{\left(\sqrt{\|\pmb{\mu}_j\|^2+1}-\bm{y}^\top_0\pmb{\mu}_j\right)}, \ \ j=1,\ldots,J
\end{eqnarray*}
and allocate $\bm{y}_0$ to the group $j$ that minimizes the above quantity. In practice, these parameters are unknown, so they are estimated, separately, for each group. In the case of the PKB distribution the allocation rule is straightforward to write down analytically, but evidently, in a more complicated form. In this case,  $\bm{y}_0$ is allocated to the group $j$ with the minimum value in the following quantity
\begin{eqnarray*} 
\log{\sqrt{\|\pmb{\mu}_j\|^2+1}-{\bf z}^\top\pmb{\mu}_j} + \log{\sqrt{\|\pmb{\mu}_j\|^2+1}-1} - \log{\|\pmb{\mu}_j\|^2}, \ \ j=1,\ldots,J.
\end{eqnarray*}

\subsection{Model based clustering}\label{sec:mix}
In this section we are dealing with the problem of discovering latent groups (clusters) in possibly heterogeneous directional datasets\footnote{The discussion covers both distributions under study and this is why it describes everything in a  generic fashion}. Under a model-based clustering point of view, we are modelling the observed data as a finite mixture of (spherical Cauchy or Poisson kernel-based) distributions. The EM algorithm is a standard approach in estimating finite mixture models \citep{mclachlan2000finite, mcnicholas2016mixture, fruhwirth2019handbook, yao2024mixture}. 

Assume for instance that there are $K\geqslant 1$ underlying clusters in the population with (unknown) weights equal to $p_1,\ldots,p_K$, where $p_j>0$ and $\sum_{j=1}^{K}p_j=1$. Define the (latent) allocation variables $\bm{Z}_i = (Z_{i1},\ldots,Z_{in})^\top$, $i=1,\ldots,n$, with $\bm{Z}_i \sim\mathcal M(1;p_1,\ldots,p_K)$, where the latter represents the multinomial distribution with $K$ categories and success probabilities equal to $p_1,\ldots,p_K$. The prior probability of selecting an observation from cluster $j$ is equal to $\mathrm{P}(Z_{ij} = 1) = p_j$ independent for  $i = 1,\ldots,n$. Conditional on $\bm{Z}_i = \bm{z}_i$, $\bm{Y}_i$ is distributed as
\begin{equation}
    \bm{Y}_i| (\bm{Z}_{i} = \bm{z}_i) \sim f_{\bm{Y}_i| \bm{Z}_{i}}(\bm{y}_i|\bm{z}_i) =\prod_{j=1}^{K} f^{z_{ij}}(\bm{y}_i;\bm{\mu}_j),\quad\mbox{independent for}\quad i = 1,\ldots,n
\end{equation}
where $f(\cdot;\bm{\mu}_j)$ denotes the probability density function of the SC or PKB distribution with (unconstrained) parameter $\bm{\mu}_j\in\mathbb R^{d+1}$ for $j=1,\ldots,K$. Notice that the previous equation in case where the $j$-th element of $\bm{Z}_{i}$ is equal to 1, collapses to $\bm{Y}_i| (Z_{ij} = 1) \sim f(\cdot;\bm{\mu}_j)$.
It follows that the joint distribution of $\bm{Y}_i, \bm{Z}_i$ is 
\begin{equation}
\label{eq:joint}
    (\bm{Y}_i, \bm{Z}_{i}) \sim f_{\bm{Y}_i, \bm{Z}_i}(\bm{y}_i, \bm{z}_i) = \prod_{j=1}^{K}\left\{p_jf(\bm{y}_i;\bm{\mu}_j)\right\}^{z_{ij}}.
\end{equation}
Consequently, the marginal distribution of $\bm{Y}_i$ is a mixture of $K$ distributions
\begin{equation}
\label{eq:fmm}
    \bm{Y}_i \sim f_{\bm{Y}_i}(\bm{Y}_i) =\sum_{j=1}^{K}p_j f(\bm{Y}_i;\bm{\mu}_j),\quad\mbox{independent for}\quad i = 1,\ldots,n.
\end{equation}
The probability that observation $i$ is assigned to component $j$, conditional on $\bm{Y}_i = \bm{Y}_i$, is equal to 
\begin{equation}
\label{eq:post}
\mathrm{P}(Z_{ij} = 1|\bm{Y}_i = \bm{y}_i) =     \frac{p_j f(\bm{y}_i;\bm{\mu}_j)}{\sum_{\ell=1}^{K}p_\ell f(\bm{y}_i;\bm{\mu}_\ell)} =:w_{ij},\quad j=1,\ldots,K
\end{equation}
or equivalently, $\bm{Z}_i|(\bm{Y}_i = \bm{Y}_i)\sim\mathcal M(1,w_{i1},\ldots,w_{iK})$, independent for $i=1,\ldots,n$. The observed log-likelihood is defined as 
\begin{equation}
\label{eq:mixlik}
\ell(\bm{\mu}, \bm{p}|\bm{y}) =  \sum_{i=1}^{n}\log\left\{\sum_{j=1}^{K}p_j f(\bm{y}_i;\bm{\mu}_j)\right\},
\end{equation}
where $\bm{\mu} = (\bm{\mu}_{1},\ldots,\bm{\mu}_K)\in\mathbb R^{K(d+1)}$ and $\bm{p} = (p_1,\ldots,p_K)\in\mathcal P_{K-1}$, where $\mathcal P_{K-1} = \{p_1,\ldots,p_K: p_j > 0, j=1,\ldots,K, \sum_{j=1}^{K}p_j = 1\}$.
In order to maximize \eqref{eq:mixlik}, the EM algorithm is implemented. In brief, given a set of starting values, the algorithm proceeds by computing the expectation of the complete log-likelihood (E-step) and then maximizing with respect to $\bm{\mu}$, $\bm{p}$ (M-step). Consider that we have at hand the \textit{complete} data $(\bm{Y}_i,\bm{Z}_i)$, independent for $i=1,\ldots,n$. From Equation \eqref{eq:joint},  the logarithm of the \textit{complete likelihood} is defined as
\begin{equation}
    \label{eq:complete_lik}
    \ell^{c}(\bm{\mu}, \bm{p}|\bm{y},\bm{z}) = \sum_{i=1}^{n}\sum_{j=1}^{K}z_{ij}\{\log p_j+\log f(y_i;\bm{\mu}_j)\}.
\end{equation}
Of course, we do not directly observe $\bm{Z}_i$, $i = 1,\ldots,n$, thus, we cannot use \eqref{eq:complete_lik}. Nevertheless, we can consider the expectation of the logarithm of the complete likelihood with respect to the conditional distribution of $\bm{Z}|\bm{y}$ and a current estimate of $\bm{\mu}$, $\bm{p}$
\begin{equation}
    \label{eq:expected_complete}
    \mathrm{E}_{\bm{Z}|\bm{y}} \ell^{c}(\bm{\mu}, \bm{p}|\bm{y},\bm{Z}) = \sum_{j=1}^{K}w_{\cdot j}\log p_{j} + \sum_{j=1}^{K}\sum_{i=1}^{n}w_{i j}\log f(\bm{Y}_i;\bm{\mu}_j),
\end{equation}
where $w_{\cdot j} := \sum_{i=1}^{n}w_{ij}$ as defined in \eqref{eq:post}.

In the M-step, we maximize \eqref{eq:expected_complete} with respect to the parameters $\bm{\mu}, \bm{p}$. It is easy to see that the mixing proportions are set equal to 
\[
p_j = \frac{w_{\cdot j}}{n},\quad j= 1,\ldots,K.
\]
However, the maximization of  \eqref{eq:expected_complete} is not available in closed form for $\bm{\mu}$, so a numerical approach was implemented. There are two approaches to this end, a) either use the NR or the hybrid algorithm, described in Section \ref{mle}, or b) use a numerical optimizer, such as Nelder-Mead's simplex method \citep{nelder1965}, available via the \textit{optim()} function in \textit{R}. 

Let now $\hat{\bm{p}}$, $\hat{\bm{\mu}}$ denote the estimates of mixing proportions and component-specific parameters, obtained at the last iteration of the EM algorithm. Define the corresponding estimates of the posterior membership probabilities as 
$$\hat w_{ij} = \frac{\hat p_j f(\bm{y}_i;\hat{\bm{\mu}}_j)}{\sum_{\ell=1}^{K}\hat{p}_\ell f(\bm{y}_i;\hat{\bm{\mu}}_\ell)},\quad j=1,\ldots,K.
$$
Then, the resulting single best clustering $\{c_1,\ldots,c_n\}$ arises after applying the Maximum A Posteriori (MAP) rule in the estimated posterior membership probabilities, that is,
\begin{equation}
    \label{eq:map}
    c_i = \mathrm{argmax}_{j}\{\hat w_{ij}, j=1\ldots,K\},\quad i = 1,\ldots,n.
\end{equation}

Initialization of the EM algorithm demands extra care  in order to avoid convergence to local maxima. Some indicative works on this topic include \cite{biernacki2003choosing, karlis2003choosing, fraley2005incremental, baudry2015mixtures, papastamoulis2016estimation, michael2016effective}. In our implementation of the mixtures of SC distributions, the suggested initialization is based on $k$-means starting values since we observed that for this particular class of finite mixture models this strategy gave  comparable results to more sophisticated initialization schemes (see Section 2.3 of \cite{papastamoulis2023model}). For the case of the PKB we relied on the implementation of \cite{golzy2020}. 

In order to select the optimal number of clusters we used information criteria such as the Bayesian Information Criterion (BIC) \citep{schwarz1978estimating} and/or Integrated Complete Likelihood (ICL) \citep{biernacki2000assessing}.

\section{Simulation studies} \label{simulations}
Simulation studies were performed to assess the computational efficiency of the two algorithms employed for MLE of the two distributions. Furthermore, a comparative analysis of the SC distribution and the PKB distribution was conducted using both spherical and hyper-spherical data. This comparison was framed within the previously discussed contexts, including the equality of two location parameters, as well as regression and discriminant analysis settings. The computations took place using \textit{R} and all algorithms are available via the package \textit{Directional} \citep{directional2024}.

\subsection{Computational efficiency of the MLE algorithms}
We compared the runtime of two MLE algorithms—namely, the hybrid algorithm and the N algorithm—across various sample sizes and dimensionalities, using data generated from either the SC or the PKB distribution. The speed-up factor of the NR algorithm relative to the hybrid algorithm is presented in Table \ref{cost}, indicating that the NR algorithm is generally preferred, especially as sample sizes increase. However, the NR algorithm's advantage diminishes with higher dimensionalities, which is expected due to its requirement for computing the inverse of the Hessian matrix. Notably, for circular data ($d=1$), Kent and Tyler's algorithm \citep{kent1988} outperforms the NR algorithm in terms of speed\footnote{Comparisons with \textit{R}'s built-in \textit{optim()} function are omitted as they are deemed unnecessary for this analysis.}.

Additionally, we compared the time required to fit the SC distribution versus the PKB distribution under data generated from both distributions. As shown in Table \ref{cost2}, the MLE of the SC distribution is consistently faster than the MLE of the PKB distribution in both scenarios. These results hold true irrespective of whether the data are generated from the SC or PKB distribution.

\begin{table}[ht]
\centering
\caption{Speed-up factors of the MLE algorithms for the SC and PKB distributions for a variety of dimensions and sample sizes. The ratio of the time required by the hybrid algorithm divided by the time required by the NR algorithm. Values higher than 1 favour NR, whereas values less than 1 favour the hybrid algorithm.}
\label{cost}
\begin{tabular}{l|ccccc|ccccc}
\toprule
& \multicolumn{5}{c}{SC distribution} & \multicolumn{5}{c}{PKB distribution} \\ \midrule 
           & $d=2$ & $d=4$ & $d=6$ & $d=9$ & $d=19$ & $d=2$ & $d=4$ & $d=6$ & $d=9$ & $d=19$ \\ \midrule
$n=100$    & 0.857 & 2.126 & 0.450 & 0.423 & 0.391 & 2.729 & 2.174 & 0.530 & 0.630 & 0.267 \\ 
$n=500$    & 5.140 & 6.072 & 0.854 & 0.752 & 2.124 & 2.082 & 2.124 & 1.035 & 0.994 & 1.672 \\ 
$n=1000$   & 7.859 & 3.580 & 1.450 & 2.947 & 1.415 & 3.622 & 1.808 & 1.728 & 6.345 & 1.225 \\ 
$n=2000$   & 6.889 & 5.838 & 3.718 & 3.406 & 1.051 & 5.915 & 6.371 & 3.559 & 3.177 & 1.739 \\ 
$n=5000$   & 8.182 & 7.261 & 4.643 & 3.311 & 2.269 & 6.634 & 5.932 & 3.449 & 2.977 & 2.179 \\ 
$n=10000$  & 9.823 & 6.279 & 4.582 & 3.617 & 1.697 & 7.301 & 6.255 & 4.090 & 3.098 & 1.239 \\ 
$n=20000$  & 9.171 & 6.463 & 4.535 & 3.895 & 1.823 & 7.978 & 5.899 & 3.763 & 3.158 & 1.409 \\
\bottomrule
\end{tabular}
\end{table}

\begin{table}[ht]
\centering
\caption{Speed-up factors of the MLE algorithms for the SC and PKB distributions for a variety of dimensions and sample sizes with data generated from either distribution. The ratio of the time required to obtain the MLE of the PKB divided by the time to obtain the MLE of the SC distribution. Values higher than 1 indicate that the MLE of the SC is faster, whereas values less than 1 indicate that the MLE of the PKB is faster.}
\label{cost2}
\begin{tabular}{l|ccccc|ccccc}
\toprule
& \multicolumn{5}{c}{SC distribution} & \multicolumn{5}{c}{PKB distribution} \\ \midrule 
           & $d=2$ & $d=4$ & $d=6$ & $d=9$ & $d=19$ & $d=2$ & $d=4$ & $d=6$ & $d=9$ & $d=19$ \\ \midrule
$n=100$   & 1.357 & 1.810 & 1.577 & 1.653 & 1.648 & 1.551 & 1.560 & 1.816 & 1.501 & 1.658 \\ 
$n=500$   & 1.303 & 1.609 & 1.523 & 1.577 & 1.731 & 1.327 & 1.663 & 1.421 & 1.607 & 1.695 \\ 
$n=1000$  & 1.356 & 1.541 & 1.782 & 1.400 & 1.491 & 1.328 & 1.675 & 1.842 & 1.711 & 1.459 \\ 
$n=2000$  & 1.371 & 1.566 & 1.521 & 1.550 & 1.458 & 0.824 & 1.902 & 1.544 & 1.799 & 1.745 \\ 
$n=5000$  & 1.462 & 1.432 & 1.572 & 1.474 & 1.536 & 1.219 & 1.346 & 1.676 & 1.614 & 2.516 \\ 
$n=10000$ & 1.436 & 1.186 & 1.455 & 1.351 & 1.565 & 1.104 & 1.577 & 1.615 & 1.806 & 1.622 \\ 
$n=20000$ & 1.489 & 1.117 & 1.353 & 1.463 & 1.634 & 1.226 & 1.672 & 1.599 & 1.485 & 2.542 \\
\bottomrule
\end{tabular}
\end{table}

\subsection{Hypothesis testing for two location parameters}
According to the simulation studies conducted by \cite{tsagris2024a} for circular and spherical data, the heterogeneous approach \citep{watson1983a, watson1983b}, which does not assume equality among the concentration parameters, was identified as the optimal test in terms of size attainment. In our study, we performed a smaller-scale simulation to estimate the type I error and power of the SC log-likelihood ratio test. However, a direct comparison with the heterogeneous approach is not feasible, as the latter is designed for comparing mean directions.

Table \ref{type} provides the estimated type I error and power of the SC log-likelihood ratio test alongside those of the heterogeneous approach. These estimates are presented for data generated from both the SC and PKB distributions across various dimensions.

\begin{table}[ht]
\centering
\caption{Estimated type I error (nominal error is set to 5\%) and estimated power when the data are generated from the SC and PKB distributions with concentration parameters equal to $\rho_1=0.3$ and $\rho_2=0.8$, for the first and second sample, respectively. The angular difference between the two location directions is given by $\theta$.}
\label{type}
\begin{tabular}{c|rrrr|rrrr}
\toprule
      & \multicolumn{4}{c}{SC distribution} & \multicolumn{4}{c}{PKB distribution} \\ \midrule   
Sample sizes ($n_1$, $n_2$)  &  \multicolumn{4}{c}{Dimensionality ($d$)} & \multicolumn{4}{c}{Dimensionality ($d$)}  \\  \midrule
$\theta=0^{\circ}$  & 2 & 4 & 6 & 9 & 2 & 4 & 6 & 9  \\ \midrule
(50, 30)            & 0.053 & 0.048 & 0.052 & 0.049 & 0.044 & 0.055 & 0.070 & 0.054 \\ 
(70, 50)            & 0.056 & 0.050 & 0.046 & 0.056 & 0.048 & 0.055 & 0.049 & 0.058 \\ 
(100, 70)           & 0.051 & 0.050 & 0.048 & 0.054 & 0.061 & 0.052 & 0.041 & 0.05 \\  \midrule
$\theta=15^{\circ}$ & 3 & 5 & 7 & 10 & 3 & 5 & 7 & 10  \\ \midrule
(50, 30)            & 0.222 & 0.366 & 0.486 & 0.635 & 0.139 & 0.166 & 0.182 & 0.201 \\ 
(70, 50)            & 0.283 & 0.487 & 0.645 & 0.811 & 0.175 & 0.198 & 0.232 & 0.243 \\ 
(100, 70)           & 0.378 & 0.656 & 0.827 & 0.947 & 0.231 & 0.256 & 0.321 & 0.37 \\  \midrule
$\theta=30^{\circ}$ & 3 & 5 & 7 & 10 & 3 & 5 & 7 & 10  \\ \midrule          
(50, 30)            & 0.633 & 0.916 & 0.984 & 0.998 & 0.404 & 0.502 & 0.558 & 0.676 \\ 
(70, 50)            & 0.798 & 0.983 & 0.999 & 1.000 &  0.528 & 0.637 & 0.738 & 0.836 \\ 
(100, 70)           & 0.914 & 1.000 & 1.000 & 1.000 & 0.690 & 0.826 & 0.885 & 0.951 \\ \bottomrule
\end{tabular}
\end{table}

\subsection{Regression analysis} 
For the regression analysis, we adhered to the methodology outlined by \cite{tsagris2024b}. We utilized a single covariate to generate $n$ values from a standard normal distribution, linking it to the response directional variable in a linear manner, $\pmb{\mu}_i={\bf X}_i{\bf B}$, for $i=1,\ldots,n$. Subsequently, data were generated from either the SC or the PKB distribution, with the median and mean directions, respectively, being ${\bf m}_i=\pmb{\mu}_i/\|\pmb{\mu}_i\|$. For both distributions, the concentration parameter was equal to $\rho_i=\left(\sqrt{\|\pmb{\mu}_i\|^2 + 1} - 1 \right) / \|\pmb{\mu}_i\|$.

This procedure was iterated for various combinations of sample sizes and dimensionalities, with regression coefficients being estimated using both the SC and PKB regression models. The entire process was repeated 1,000 times, and the average fit of the two models, as measured by the quantity $\sum_{i=1}^n\bm{y}_i^\top\hat{\bm{y}}_i/n$ is presented in Table \ref{fit}. The fit takes values from 0 up to 1, where higher values indicate better fit. The estimated fits of the regression models are nearly nearly identical.

\begin{table}[ht]
\centering
\caption{Estimated fit of each regression model, computed as $\sum_{i=1}^n\bm{y}_i^\top\hat{\bm{y}}_i/n$ when the data are generated from the SC and the PKB distributions.}
\label{fit}
\begin{tabular}{c|l|rrrr|rrrr}
  \toprule
          &  & \multicolumn{4}{c}{SC distribution}  &  \multicolumn{4}{c}{PKB distribution}  \\ \midrule
          &  & \multicolumn{4}{c}{Dimensionality ($d$)} & \multicolumn{4}{c}{Dimensionality ($d$)} \\ \midrule
Sample size ($n$)  & Model  & 2 & 4 & 6 & 9 & 2 & 4 & 6 & 9 \\ \midrule
50  & SC  & 0.473 & 0.851 & 0.895 & 0.965 & 0.584 & 0.666 & 0.783 & 0.799 \\ 
    & PKB & 0.472 & 0.851 & 0.895 & 0.965 & 0.584 & 0.666 & 0.782 & 0.798 \\ \midrule
100 & SC  & 0.755 & 0.850 & 0.945 & 0.967 & 0.491 & 0.595 & 0.710 & 0.779 \\ 
    & PKB & 0.754 & 0.850 & 0.945 & 0.967 & 0.491 & 0.595 & 0.710 & 0.778 \\ \midrule
200 & SC  & 0.769 & 0.910 & 0.927 & 0.972 & 0.532 & 0.647 & 0.689 & 0.759 \\ 
    & PKB & 0.769 & 0.910 & 0.927 & 0.972 & 0.532 & 0.647 & 0.689 & 0.759 \\ \bottomrule
\end{tabular}
\end{table}

\subsection{Discriminant analysis}
Following the approach of \cite{tsagris2019}, we also simulated data from the SC and PKB distributions, assuming two groups with location parameters differing by a specific angle. A 10-fold cross-validation protocol was employed to estimate the percentage of correct classification\footnote{A notable difference from \cite{tsagris2019} is that our simulations were extended to higher-dimensional settings.}. This procedure was repeated 1,000 times, and the average classification percentages are reported in Table \ref{discrim}. It is evident that, irrespective of the underlying data generation mechanism, the classification performance of the two distributions is nearly identical.

\begin{table}[ht]
\centering
\caption{Estimated percentage of correct classification when the data are generated from the SC and the PKB distributions, and the concentration parameter was equal to $\rho=0.5$. The angle between the location parameters of the two groups is denoted by $\theta$.}
\label{discrim}
\begin{tabular}{c|r|rrrr|rrrr}
\toprule
      & & \multicolumn{4}{c}{SC distribution} & \multicolumn{4}{c}{PKB distribution} \\ \midrule   
Sample size ($n$)  & &  \multicolumn{4}{c}{Dimensionality ($d$)} & \multicolumn{4}{c}{Dimensionality ($d$)}  \\  \midrule
$\theta=15^{\circ}$ & Model & 2 & 4 & 6 & 9 & 2 & 4 & 6 & 9  \\ \midrule
50  & SC  & 0.560 & 0.603 & 0.628 & 0.664 & 0.540 & 0.543 & 0.550 & 0.561 \\ 
    & PKB & 0.561 & 0.603 & 0.630 & 0.665 & 0.540 & 0.542 & 0.551 & 0.562 \\ \hline
100 & SC  & 0.575 & 0.617 & 0.639 & 0.676 & 0.543 & 0.557 & 0.563 & 0.573 \\ 
    & PKB & 0.576 & 0.618 & 0.641 & 0.677 & 0.545 & 0.558 & 0.564 & 0.574 \\ \hline
200 & SC  & 0.581 & 0.623 & 0.649 & 0.684 & 0.553 & 0.564 & 0.573 & 0.582 \\ 
    & PKB & 0.581 & 0.623 & 0.649 & 0.685 & 0.554 & 0.565 & 0.574 & 0.582 \\  \midrule
$\theta=30^{\circ}$ & Model & 2 & 4 & 6 & 9 & 2 & 4 & 6 & 9  \\ \midrule          
50  & SC  & 0.652 & 0.722 & 0.771 & 0.820 & 0.599 & 0.621 & 0.646 & 0.657 \\ 
    & PKB & 0.653 & 0.724 & 0.773 & 0.821 & 0.599 & 0.623 & 0.648 & 0.660 \\ \hline
100 & SC  & 0.658 & 0.729 & 0.779 & 0.832 & 0.612 & 0.630 & 0.651 & 0.672 \\ 
    & PKB & 0.658 & 0.730 & 0.778 & 0.832 & 0.612 & 0.630 & 0.652 & 0.673 \\ \hline
200 & SC  & 0.660 & 0.735 & 0.782 & 0.832 & 0.616 & 0.638 & 0.658 & 0.681 \\ 
    & PKB & 0.660 & 0.735 & 0.782 & 0.833 & 0.617 & 0.639 & 0.659 & 0.681 \\  \bottomrule
\end{tabular}
\end{table}

\subsection{Model based clustering}
Simulated synthetic data from \eqref{eq:fmm} as follows. The number of clusters varied on the set $g\in\{2,\ldots,5\}$. Conditional on $K$, mixing proportions were generated $p\sim\mathcal D(5,\ldots,5)$, where $\mathcal D(\alpha_1,\ldots,\alpha_j)$ denotes the $d$-dimensional Dirichlet distribution with concentration parameter $(\alpha_1,\ldots,\alpha_K)$, where $\alpha_j>0$, for all $j$. The location direction $\bf m_j$ and concentration parameter $\rho_j$ for cluster $j$ were drawn from an arbitrary von Mises-Fisher distribution and simulated uniformly from $(0.7,0.9)$, respectively. 

For each combination of number of clusters and dimensionality 200 replicates of the previously described mechanism were considered. The sample size was set equal to $n = 500$ and $n = 1000$. For each case, finite mixture models of the form \eqref{eq:fmm} were fitted using the EM algorithm\footnote{For the the mixtures of PKB distributions we used the package \textit{QuadratiK} \cite{quadratik2024}} considering that possible values for the number of clusters were $K = 1,\ldots,K_{\mathrm{max}}$ where $K_{\mathrm{max}} = K+3$. The optimal number of clusters was selected according to BIC, but the adjusted Rand index (ARI) is also reported. 

Table \ref{bic} presents the mean absolute difference between the true and the estimated number of cluster $|\hat{K}-K|/200$. When the true mixture model is the SC distribution the mixtures of SC perform worse than the mixtures of PKB distributions when $d=2$ and $d=4$ and better when $d=6$ and $d=9$, regardless of the sample size. When the true mixture model is the PKB distribution, the results change slightly. When $d=2$, the mixtures of PKB distributions perform better, for both sample sizes. However, when $d=4$, the PKB performs worse than SC for $n=500$, but better when $n=1000$. Finally, when $d=6$ and $d=9$ the mixtures of PKB distributions are worse than the mixtures of SC distribution, regardless of the sample size.

Table \ref{ari} contains the average ARI values. When the data are generated from the SC distribution, the average ARI values are high for both models and increase with increasing dimensions. On the contrary, when the data are generated from the PKB distribution, the mixtures of PKB distributions produce higher ARI values and once again they increase with increasing dimensionalities.  

\setlength{\tabcolsep}{2.5pt}
\begin{table}[ht]
\centering
\begin{small}
\caption{ Estimate of the absolute error $|\hat{K}-K|$ where $\hat K$ and $K$ denote the estimated and true number of clusters, respectively. We considered various combinations of dimensionalities ($d$), true number of clusters ($K$) and sample size $n$. In all cases the data are generated from the SC and the PKB distributions.}
\label{bic}
\begin{tabular}{r|rr|rr|rr|rr|rr|rr|rr|rr}
\toprule
& \multicolumn{8}{c}{SC distribution} & \multicolumn{8}{c}{PKB distribution} \\ \midrule
 & \multicolumn{2}{c}{$d=2$} & \multicolumn{2}{c}{$d=4$} & \multicolumn{2}{c}{$d=6$} & \multicolumn{2}{c}{$d=9$}
& \multicolumn{2}{c}{$d=2$} & \multicolumn{2}{c}{$d=4$} & \multicolumn{2}{c}{$d=6$} & \multicolumn{2}{c}{$d=9$} \\ \midrule
$K$ & SC & PKB & SC & PKB & SC & PKB & SC & PKB & SC & PKB & SC & PKB & SC & PKB & SC & PKB \\ \midrule
($n=500$) 2 & 0.240 & 0.030 & 0.320 & 0.185 & 0.280 & 0.565 & 0.055 & 1.120 & 0.380 & 0.050 & 0.885 & 1.395 & 1.160 & 2.345 & 1.230 & 2.450 \\ 
3 & 0.285 & 0.075 & 0.475 & 0.235 & 0.345 & 0.925 & 0.085 & 1.560 & 0.400 & 0.155 & 0.880 & 1.710 & 1.035 & 2.505 & 1.120 & 2.660 \\ 
4 & 0.400 & 0.145 & 0.555 & 0.275 & 0.450 & 1.090 & 0.150 & 1.555 & 0.600 & 0.425 & 0.850 & 1.830 & 0.945 & 2.550 & 0.885 & 2.815 \\ 
5 & 0.590 & 0.400 & 0.625 & 0.540 & 0.435 & 1.365 & 0.230 & 1.720 & 0.970 & 0.770 & 0.745 & 1.955 & 1.075 & 2.665 & 0.795 & 2.820 \\  \midrule
($n=1000$) 2 & 0.040 & 0.035 & 0.145 & 0.045 & 0.150 & 0.165 & 0.055 & 0.390 & 0.815 & 0.245 & 1.325 & 0.535 & 1.455 & 1.565 & 1.645 & 2.170 \\ 
3 & 0.110 & 0.065 & 0.170 & 0.035 & 0.110 & 0.325 & 0.115 & 0.645 & 0.620 & 0.255 & 1.245 & 0.640 & 1.610 & 1.890 & 1.700 & 2.225 \\ 
4 & 0.265 & 0.110 & 0.260 & 0.060 & 0.220 & 0.325 & 0.080 & 0.895 & 0.595 & 0.355 & 1.155 & 0.600 & 1.370 & 2.065 & 1.635 & 2.435 \\ 
5 & 0.395 & 0.255 & 0.315 & 0.125 & 0.305 & 0.555 & 0.050 & 0.970 & 0.885 & 0.685 & 1.190 & 0.765 & 1.280 & 2.095 & 1.440 & 2.485 \\ 
 \bottomrule
\end{tabular}
\end{small}
\end{table}

\begin{table}[ht]
\centering
\begin{small}
\caption{Average ARI for various combinations of dimensionalities ($d$), true number of clusters ($K$) and sample size $n$. In all cases the data are generated from the SC and the PKB distributions.}
\label{ari}
\begin{tabular}{r|rr|rr|rr|rr|rr|rr|rr|rr}
\toprule
& \multicolumn{8}{c}{SC distribution} & \multicolumn{8}{c}{PKB distribution} \\ \midrule
& \multicolumn{2}{c}{$d=2$} & \multicolumn{2}{c}{$d=4$} & \multicolumn{2}{c}{$d=6$} & \multicolumn{2}{c}{$d=9$}
& \multicolumn{2}{c}{$d=2$} & \multicolumn{2}{c}{$d=4$} & \multicolumn{2}{c}{$d=6$} & \multicolumn{2}{c}{$d=9$} \\ \midrule
$K$ & SC & PKBD & SC & PKBD & SC & PKBD & SC & PKBD & SC & PKBD & SC & PKBD & SC & PKBD & SC & PKBD \\ \midrule
($n=500$) 2 & 0.818 & 0.818 & 0.968 & 0.972 & 0.991 & 0.989 & 0.997 & 0.985 & 0.603 & 0.645 & 0.594 & 0.777 & 0.586 & 0.843 & 0.586 & 0.910 \\ 
3 & 0.778 & 0.775 & 0.957 & 0.969 & 0.988 & 0.991 & 0.994 & 0.987 & 0.570 & 0.601 & 0.644 & 0.753 & 0.664 & 0.823 & 0.719 & 0.902 \\ 
4 & 0.773 & 0.769 & 0.953 & 0.960 & 0.980 & 0.990 & 0.997 & 0.989 & 0.503 & 0.525 & 0.614 & 0.700 & 0.691 & 0.801 & 0.772 & 0.882 \\ 
5 & 0.697 & 0.698 & 0.950 & 0.955 & 0.979 & 0.987 & 0.993 & 0.990 & 0.471 & 0.499 & 0.608 & 0.684 & 0.662 & 0.762 & 0.779 & 0.868 \\  \midrule
($n=1000$)2 & 0.837 & 0.834 & 0.978 & 0.969 & 0.988 & 0.997 & 0.998 & 0.997 & 0.558 & 0.641 & 0.539 & 0.786 & 0.521 & 0.852 & 0.520 & 0.914 \\ 
3 & 0.772 & 0.768 & 0.968 & 0.965 & 0.993 & 0.994 & 0.995 & 0.998 & 0.540 & 0.581 & 0.611 & 0.759 & 0.617 & 0.833 & 0.649 & 0.900 \\ 
4 & 0.745 & 0.744 & 0.966 & 0.964 & 0.990 & 0.994 & 0.998 & 0.991 & 0.516 & 0.541 & 0.598 & 0.704 & 0.656 & 0.807 & 0.696 & 0.883 \\ 
5 & 0.717 & 0.715 & 0.954 & 0.951 & 0.988 & 0.984 & 0.999 & 0.995 & 0.469 & 0.500 & 0.580 & 0.681 & 0.646 & 0.771 & 0.726 & 0.869 \\  \bottomrule
\end{tabular}
\end{small}
\end{table}

\section{Real data analysis} \label{real}

\subsection{Hypothesis testing for two location parameters}
\begin{figure}[!ht]
\centering
\includegraphics[scale = 0.65]{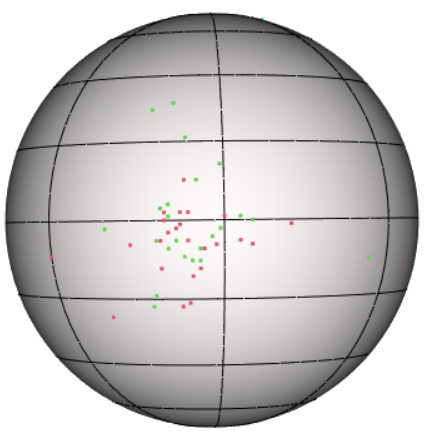} 
\caption{Ordovician data. The red and green indicate the two groups of the data.}
\label{sphere}
\end{figure}

The Ordovician dataset \citep{fisher1993} comprises two groups, each containing 50 measurements on the sphere, derived from $L_0^1$ axes (intersections between cleavage and bedding planes of F-folds) in Ordovician turbidites, collected within the same sub-domain. Figure \ref{sphere} illustrates the dataset on the sphere, with colors distinguishing the two groups. Table \ref{ordo} presents the MLE for both the SC and PKB distributions. As shown, the differences between the two models are minimal. The log-likelihood ratio tests for both the SC and PKB models yielded high p-values, 0.733 and 0.856, respectively. Additionally, their corresponding bootstrap-based p-values were similarly high, at 0.825 and 0.914 for the SC and PKB models, respectively.

\begin{table}[ht]
\centering
\caption{Estimated parameters of the SC and PKB for the Ordovician data.}
\label{ordo}
\begin{tabular}{r|rrr|r|rrr|r}
\toprule
& \multicolumn{4}{c}{SC distribution} & \multicolumn{4}{c}{PKB distribution} \\ \midrule
& \multicolumn{3}{c}{$\hat{\bf m}$} & $\hat{\rho}$ & \multicolumn{3}{c}{$\hat{\bf m}$} & $\hat{\rho}$ \\ \midrule 
Group 1 & 0.770 & 0.635 & -0.067 & 0.853 & 0.778 & 0.626 & -0.050 & 0.908 \\ 
Group 2 & 0.761 & 0.648 & -0.035 & 0.812 & 0.764 & 0.644 & -0.048 & 0.875 \\ 
\bottomrule
\end{tabular}
\end{table}

\subsection{Regression analysis}
Data concerning crop productivity in the Greek NUTS II region of Thessaly during the 2017–2018 cropping year were provided by the Greek Ministry of Agriculture, specifically through the farm accountancy data network (FADN). The dataset pertains to a sample of 487 farms and initially included 20 crops. However, after aggregation, the number of crops was reduced to 10\footnote{A larger version of this dataset was employed in \cite{mattas2024}.}. For each of the 487 farms, both the cultivated area and the production for each of the 10 crops are recorded. The objective of this study is to relate the composition of production (simplicial response, $\mathbf{Y}$) to the composition of cultivated area (simplicial predictor, $\mathbf{X}$). Consequently, the data were scaled to sum to unity\footnote{The raw data cannot be distributed due to disclosure restrictions.}. A square root transformation was applied to the compositional data (both the response and predictor variables), mapping them onto a 10-dimensional sphere.

In total, including the constant terms, 110 regression parameters needed to be estimated. The SC-based regression model completed in 0.11 seconds, while the PKB regression model encountered numerical instability issues with the Hessian matrix. As a result, a numerical optimizer (the function \textit{optim()} in \textit{R}) was used, which required over 5 minutes to complete. This is unsurprising, as such optimizers are not typically suited to high-dimensional problems.

Regarding model performance, the fit, measured by $\sum_{i=1}^n \mathbf{y}_i^\top \hat{\mathbf{y}}_i / n$, was 0.958 for the SC regression model and 0.955 for the PKB regression model. Due to the significant computational time required by the PKB model, the 10-fold cross-validation procedure was not conducted.

\subsection{Discriminant analysis}
For this task, we utilize the \textit{Wireless Indoor Localization} dataset, which is publicly available on the \href{https://archive.ics.uci.edu/dataset/422/wireless+indoor+localization}{UCI Machine Learning Repository} website. The data were collected in an indoor space by recording signal strengths from seven WiFi signals detectable by a smartphone. The dataset comprises 2,000 measurements across 7 variables, representing WiFi signal strengths received from 7 routers in an office building in Pittsburgh, USA. The grouping variable corresponds to one of four rooms, with 500 observations per room. WiFi signal strength is measured in decibel milliwatts (dBm), expressed as a negative value ranging from -100 to 0. In preparation for the discriminant analysis, the data were first normalized by projecting them onto the hyper-sphere. 

\begin{figure}[!ht]
\centering
\includegraphics[scale = 0.4]{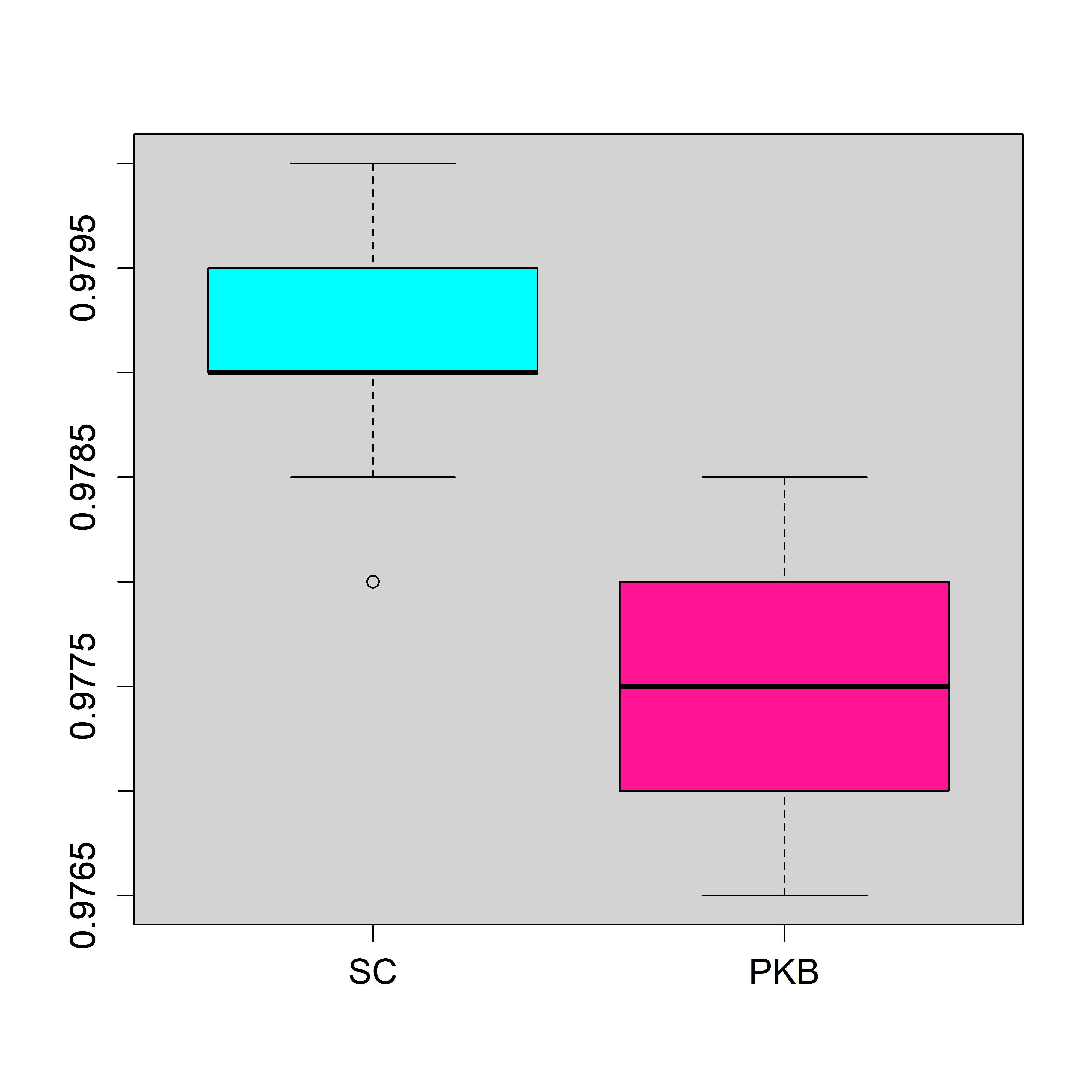}
\caption{Boxplots of the percentage of correct classification for the SC and PKB likelihood discriminant analysis applied to the \textit{Wireless Indoor Localization} data.}
\label{real_discrim}
\end{figure}

We then applied a 10-fold cross-validation procedure, repeated 50 times to account for variability due to different data splits, and computed the percentage of correct classifications at each repetition. The results are displayed in Figure \ref{real_discrim}. The average and median percentages of correct classification were 0.9792 and 0.9790, respectively, for the SC distribution, and equal to 0.9775 and 0.9775, respectively, for the PKB distribution.  

\subsection{Model based clustering}
We utilize the \textit{Wireless Indoor Localization} dataset again, on the grounds that it is a hard case, as pointed out by \cite{golzy2020} as well.  We searched for the optimal number of components (clusters) using the BIC \citep{schwarz1978estimating} and ICL \citep{biernacki2000assessing} criteria. 

We fitted mixtures of SC distributions with $1\leqslant K \leqslant 10$ components. The EM algorithm was initialized under a $k$-means strategy. For each model we applied the MAP rule \eqref{eq:map} to the estimated posterior membership probabilities and calculated the Adjusted Rand index between the estimated clustering and the ground-truth classification, that is, the room where each router is located. The results are reported in Table \ref{tab:ari}. Observe that the Adjusted Rand index is maximized when the number of mixture components is equal to 4, that is, equal to the number of rooms, corroborating the results of \cite{golzy2020}. 

\begin{figure}
\begin{tabular}{cc}
\includegraphics[width=0.45\linewidth]{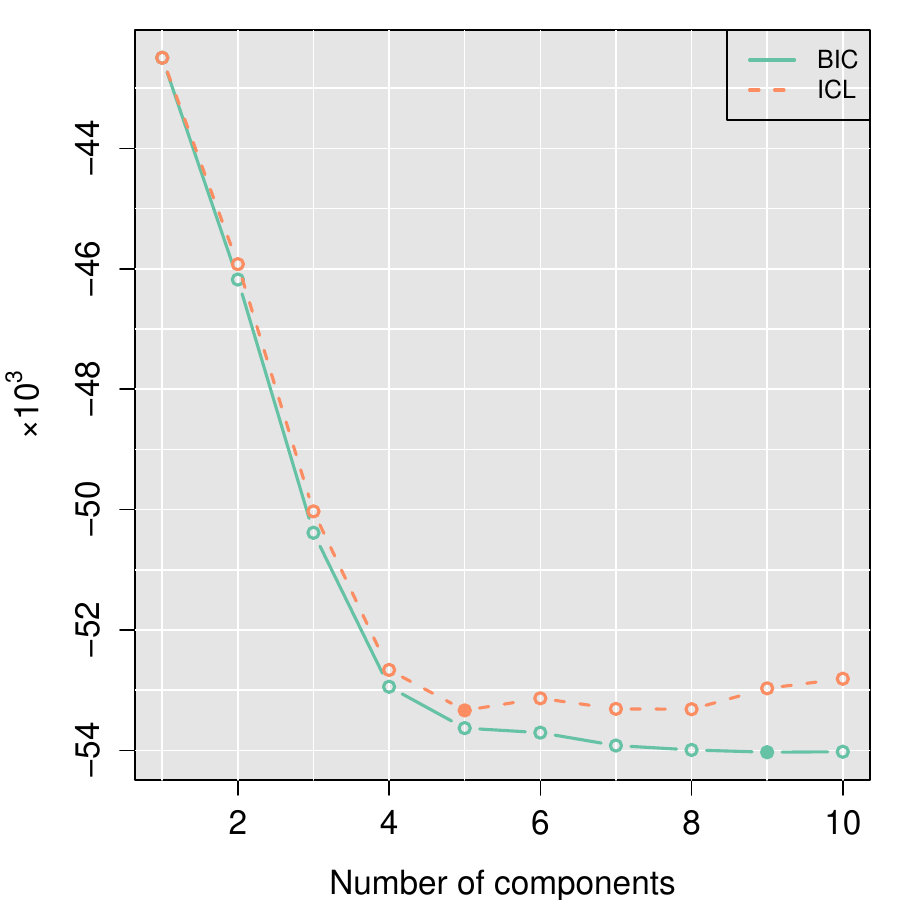}  &  
\includegraphics[width=0.45\linewidth]{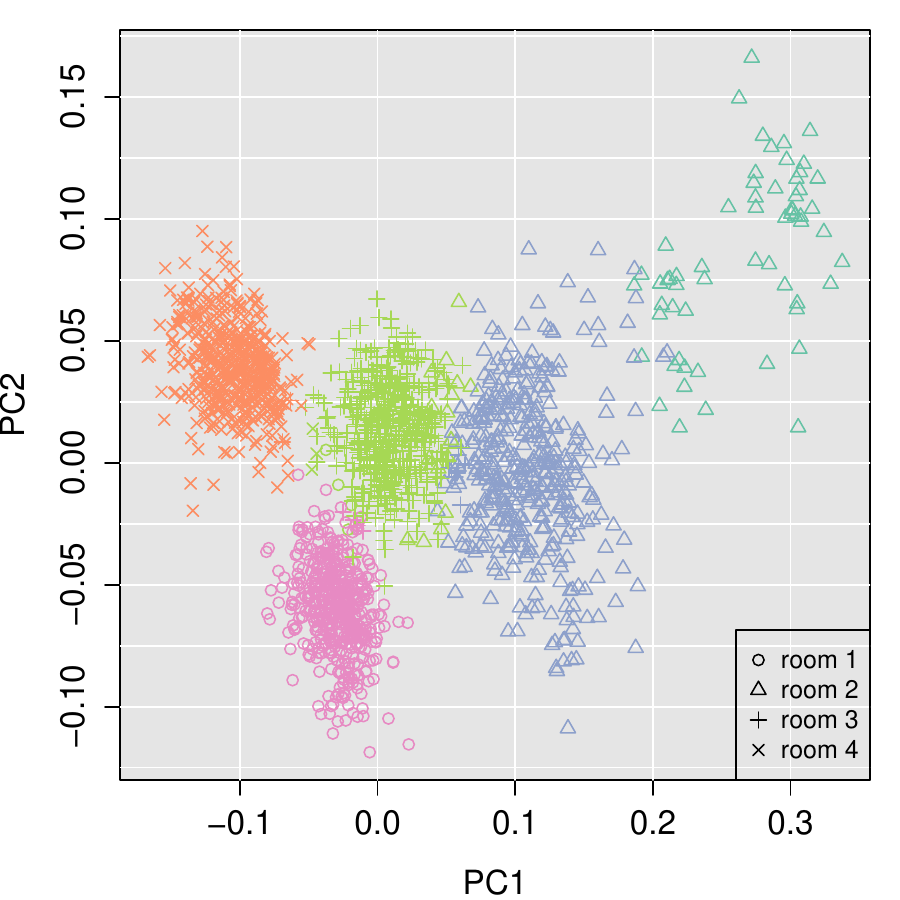} \\
(a) & (b)
\end{tabular}
\caption{Wireless data clustering: (a) BIC and ICL values for mixtures of Spherical Cauchy distributions with number of components between 1 and 10 and (b) projection of the data to the first two principal components  and coloured according to the clustering arising from a mixture model with $K=5$ Spherical Cauchy components (selected by ICL). A distinct symbol is used to denote the room where each router is located.}
\label{fig:wirelss_cluster}
\end{figure}

However, the information criteria suggest that a larger numbers of components is required, as shown in Figure \ref{fig:wirelss_cluster}(a): in particular ICL selects a model with $K=5$ while BIC selects a model with $K=9$ components. Notice that  models with $K \geqslant 5$ have very small differences in the values of the aforementioned criteria, making it hard to select one of them. In Figure \ref{fig:wirelss_cluster}(b) the resulting clusters when $K=5$ are displayed with different colour, after projecting the data into the first two principal components (explained variability $86.16\%$). Observe that the orange, purple and green clusters are actually corresponding to rooms 1, 3 and 4, respectively. However, the observations from room 2 (triangles) are split into two distinct clusters (the blue and green ones). This is somewhat expected, since the projection onto the first two principle components reveals that the observations from room 2 exhibit heterogeneity, therefore the selection of at least 5 clusters for this dataset is justified.  

We also fitted mixtures of PKB distributions trying the same number of components. Based on BIC, the optimal number of components was equal to 10, while based on the ICL 6 components were selected. 

\begin{table}[ht]
\centering
\begin{tabular}{rrrrrrrrrrr}
  \hline
 & 1 & 2 & 3 & 4 & 5 & 6 & 7 & 8 & 9 & 10 \\ 
  \hline
ARI & 0.00 & 0.31 & 0.65 & 0.94 & 0.91 & 0.85 & 0.82 & 0.79 & 0.70 & 0.67 \\ 
   \hline
\end{tabular}
\label{tab:ari}
\caption{ARI between the ground truth classification (rooms) of the Wireless Indoor Localization dataset and the estimated clusters arising from the EM algorithm when fitting mixtures of Spherical Cauchy distributions with number of components between 1 and 10.}
\end{table}

\section{Conclusions} \label{conclusions}
We investigated two recently proposed distributions, namely the spherical Cauchy (SC) and the Poisson kernel-based (PKB) distributions. We introduced an alternative, hybrid method for estimating the parameters of both distributions. For the SC distribution, we proposed a new parameterization that facilitates the application of the Newton-Raphson algorithm. Additionally, we re-parameterized the density function of the PKB distribution in a similar manner to the SC distribution. However, this re-parameterization did not prove useful for maximum likelihood estimation (MLE) of the PKB parameters. Nevertheless, the new parameterizations enabled the implementation of regression models by circumventing the (hyper-)spherical constraint, thus allowing for heterosceadstic errors similar to those found in the ESAG distribution \citep{paine2020}. The advantage of the re-parameterized SC distribution is that the Newton-Raphson method could also be applied in the regression context, leading to computationally efficient estimation of the regression parameters. The hybrid estimation method for both distributions further enabled hypothesis testing of location directions between two populations using the log-likelihood ratio test, without the assumption of equal concentration parameters. Lastly, we explored classification as well as model-based clustering under a maximum likelihood framework using either distribution.

The results of simulation studies and real data applications demonstrated the comparable performance of both distributions. However, the SC distribution stands out as the more practical choice for researchers and practitioners due to its ease of simulation, faster parameter estimation—both with and without covariates—and overall computational efficiency.

Regarding future work, several directions are noteworthy. Extending hypothesis testing to more than two location parameters is straightforward, though the primary challenge lies in computational time, which increases with the number of groups. Secondly, rejection sampling for the PKB using the SC distribution as an envelope function did not yield satisfactory results. Although we matched the SC distribution parameters to those of the PKB distribution and estimated the bound between the ratio of the two distributions, this bound increases with dimensionality, and the accuracy of this method appears inferior to the rejection sampling approach proposed by \cite{sablica2023}. A log-likelihood ratio test based rotational symmetry test was explored but the simulation studies showed that it was inferior to the Rayleigh uniformity test \citep{rayleigh1919,jupp2001}. 

\clearpage
\section*{Appendix}
\setcounter{section}{0}
\renewcommand{\thesubsection}{A\arabic{subsection}}
\setcounter{equation}{0}
\renewcommand{\theequation}{\thesubsection.\arabic{equation}}

\subsection{Difference in the log densities of the SC and PKB distributions}
The difference of the log-densities, of PKB and SC, can be written as
\begin{eqnarray*}
\log{f_{SC}}-\log{f_{PKB}} &=& \frac{d-1}{2}\left[ -\log{\left(\sqrt{\gamma^2+1}-\alpha\right)} + \log{\left(1-\frac{(\sqrt{\gamma^2+1}-1)^2}{\gamma^2}\right)}\right] \\
&=& \frac{d-1}{2}\left[\log{\frac{1+\rho^2-2\bm{y}^\top{\bf m}}{1-\rho^2}}+\log{\left(1-\rho^2\right)} \right] \\
&=& \frac{d-1}{2}\log{\left(1+\rho^2-2\bm{y}^\top{\bf m}\right)}.
\end{eqnarray*}

\subsection{Derivatives of the log-likelihood of SC, Eq. (\ref{lik2})}
\begin{eqnarray*}
{\bf J} &=& \frac{\partial\ell_{SC}}{\partial \pmb{\mu}} = - d\sum_{i=1}^n\dfrac{\frac{\pmb{\mu}}{\sqrt{\gamma^2+1}}-\bm{y}_i}{\sqrt{\gamma^2+1}-\alpha_i} \\
{\bf H} &=& \frac{\partial^2\ell_{SC}}{\partial \pmb{\mu} \partial\pmb{\mu}^\top} = - d\sum_{i=1}^n\dfrac{\frac{{\bf I}_{d+1}\sqrt{\gamma^2+1}-\frac{\pmb{\mu}\pmb{\mu}^\top}{\sqrt{\gamma^2+1}}}{\gamma^2+1}\left(\sqrt{\gamma^2+1}-\alpha_i\right)-\left(\frac{\pmb{\mu}}{\sqrt{\gamma^2+1}}-\bm{y}_i\right)\left(\frac{\pmb{\mu}}{\sqrt{\gamma^2+1}}-\bm{y}_i\right)^\top}{\left(\sqrt{\gamma^2+1}-\alpha_i\right)^2}.
\end{eqnarray*}

\subsection{Derivatives of the log-likelihood of PKB, Eq. (\ref{pkblik2})}
\begin{eqnarray*}
{\bf J} = \frac{\partial\ell_{PKB}}{\partial \pmb{\mu}} &=& - \frac{d+1}{2}\sum_{i=1}^n\dfrac{\frac{\pmb{\mu}}{\sqrt{\gamma^2+1}}-\bm{y}_i}{\sqrt{\gamma^2+1}-\alpha_i} - n\frac{d-1}{2}\left( \dfrac{\frac{\pmb{\mu}}{\sqrt{\gamma^2+1}}}{\sqrt{\gamma^2+1}-1} -\frac{2\pmb{\mu}}{\gamma^2}\right)
\\
{\bf H} = \frac{\partial^2\ell_{PKB}}{\partial \pmb{\mu} \partial\pmb{\mu}^\top} &=& - \frac{d+1}{2}\sum_{i=1}^n\dfrac{\frac{{\bf I}_{d+1}\sqrt{\gamma^2+1}-\frac{\pmb{\mu}\pmb{\mu}^\top}{\sqrt{\gamma^2+1}}}{\gamma^2+1}\left(\sqrt{\gamma^2+1}-\alpha_i\right)-\left(\frac{\pmb{\mu}}{\sqrt{\gamma^2+1}}-\bm{y}_i\right)\left(\frac{\pmb{\mu}}{\sqrt{\gamma^2+1}}-\bm{y}_i\right)^\top}{\left(\sqrt{\gamma^2+1}-\alpha_i\right)^2} \\
&& - n\frac{d-1}{2}\left[\dfrac{\frac{{\bf I}_{d+1}\sqrt{\gamma^2+1}-\frac{\pmb{\mu}\pmb{\mu}^\top}{\sqrt{\gamma^2+1}}}{\gamma^2+1}\left(\sqrt{\gamma^2+1}-1\right)-\left(\frac{\pmb{\mu}}{\sqrt{\gamma^2+1}}\right)\left(\frac{\pmb{\mu}}{\sqrt{\gamma^2+1}}\right)^\top}{\left(\sqrt{\gamma^2+1}-1\right)^2} - \frac{2{\bf I}_{d+1}\gamma^2-4\pmb{\mu}\pmb{\mu}^\top}{\gamma^4}\right].
\end{eqnarray*}

\subsection{Derivatives of the log-likelihood of SC with covariates, Eq. (\ref{screg})} \label{sec:derivatives_SC}
\begin{eqnarray*}
\frac{\partial\ell_{SC}}{\partial \pmb{\beta}_k} &=& - d\sum_{i=1}^n\dfrac{\frac{\pmb{\mu}_{ik}{\bf x}_i}{\sqrt{\gamma_i^2+1}}-\bm{y}_{ik}{\bf x}_i}{\sqrt{\gamma_i^2+1}-\alpha_i} \\
\frac{\partial^2\ell_{SC}}{\partial \pmb{\beta}_k\partial\pmb{\beta}_l^\top} &=&
\left\lbrace 
\begin{array}{cc}
- d\sum_{i=1}^n\frac{\frac{{\bf x}_i{\bf x}_i^\top\sqrt{\gamma_i^2+1}-\frac{{\bf x}_i\pmb{\mu}_{ik}\pmb{\mu}_{ik}{\bf x}_i^\top}{\sqrt{\gamma_i^2+1}}}{\gamma_i^2+1}\left(\sqrt{\gamma_i^2+1}-\alpha_i\right)-\left(\frac{{\bf x}_i\pmb{\mu}_{ik}}{\sqrt{\gamma_i^2+1}}-\bm{y}_{ik}{\bf x}_i\right)\left(\frac{{\bf x}_i\pmb{\mu}_{ik}}{\sqrt{\gamma_i^2+1}}-\bm{y}_{ik}{\bf x}_i\right)^\top}{\left(\sqrt{\gamma_i^2+1}-\alpha_i\right)^2}, & \text{if} \ k=l \\
- d\sum_{i=1}^n\frac{\frac{-\frac{{\bf x}_i\pmb{\mu}_{ik}\pmb{\mu}_{il}{\bf x}_i^\top}{\sqrt{\gamma_i^2+1}}}{\gamma_i^2+1}\left(\sqrt{\gamma_i^2+1}-\alpha_i\right)-\left(\frac{{\bf x}_i\pmb{\mu}_{ik}}{\sqrt{\gamma_i^2+1}}-\bm{y}_{ik}{\bf x}_i\right)\left(\frac{{\bf x}_i\pmb{\mu}_{il}}{\sqrt{\gamma_i^2+1}}-\bm{y}_{il}{\bf x}_i\right)^\top}{\left(\sqrt{\gamma_i^2+1}-\alpha_i\right)^2}, & \text{if} \ k \neq l \\
\end{array}
\right\rbrace
\end{eqnarray*}

\subsection{Derivatives of the log-likelihood of PKB with covariates, Eq. (\ref{pkbdreg})} \label{sec:derivatives_PKB}
The vector of the first derivative is given by
\begin{eqnarray*}
\frac{\partial\ell_{PKB}}{\partial \pmb{\beta}_k} =
- \frac{d+1}{2}\sum_{i=1}^n\dfrac{\frac{\pmb{\mu}_{ik}{\bf x}_i}{\sqrt{\gamma_i^2+1}}-\bm{y}_{ik}{\bf x}_i}{\sqrt{\gamma_i^2+1}-\alpha_i} - \frac{d-1}{2}\sum_{i=1}^n\dfrac{\frac{\pmb{\mu}_{ik}{\bf x}_i}{\sqrt{\gamma_i^2+1}}}{\sqrt{\gamma_i^2+1}-1} + \frac{d-1}{2}\sum_{i=1}^n\frac{2\pmb{\mu}_{ik}{\bf x}_i}{\gamma_i^2}.
\end{eqnarray*}
The Jacobian matrix of the second derivatives comprises of
\begin{eqnarray*}
\frac{\partial^2\ell_{PKB}}{\partial \pmb{\beta}_k\partial\pmb{\beta}_k^\top} &=&
- \frac{d+1}{2}\sum_{i=1}^n\frac{\frac{{\bf x}_i{\bf x}_i^\top\sqrt{\gamma_i^2+1}-\frac{{\bf x}_i\pmb{\mu}_{ik}\pmb{\mu}_{ik}{\bf x}_i^\top}{\sqrt{\gamma_i^2+1}}}{\gamma_i^2+1}\left(\sqrt{\gamma_i^2+1}-\alpha_i\right)-\left(\frac{{\bf x}_i\pmb{\mu}_{ik}}{\sqrt{\gamma_i^2+1}}-\bm{y}_{ik}{\bf x}_i\right)\left(\frac{{\bf x}_i\pmb{\mu}_{ik}}{\sqrt{\gamma_i^2+1}}-\bm{y}_{ik}{\bf x}_i\right)^\top}{\left(\sqrt{\gamma_i^2+1}-\alpha_i\right)^2} \\ 
&& - \frac{d-1}{2}\sum_{i=1}^n\frac{\frac{{\bf x}_i{\bf x}_i^\top\sqrt{\gamma_i^2+1}-\frac{{\bf x}_i\pmb{\mu}_{ik}\pmb{\mu}_{ik}{\bf x}_i^\top}{\sqrt{\gamma_i^2+1}}}{\gamma_i^2+1}\left(\sqrt{\gamma_i^2+1}-1\right)-\left(\frac{{\bf x}_i\pmb{\mu}_{ik}}{\sqrt{\gamma_i^2+1}}\right)\left(\frac{{\bf x}_i\pmb{\mu}_{ik}}{\sqrt{\gamma_i^2+1}}\right)^\top}{\left(\sqrt{\gamma_i^2+1}-1\right)^2} \\
&& + \frac{d-1}{2}\sum_{i=1}^n\frac{2{\bf x}_i{\bf x}_i^\top\gamma_i^2-4\mu_{ik}{\bf x}_i\mu_{ik}{\bf x}_i^\top}{\gamma_i^4}, \\
\frac{\partial^2\ell_{PKB}}{\partial \pmb{\beta}_k\partial\pmb{\beta}_l^\top} &=&
- \frac{d+1}{2}\sum_{i=1}^n\frac{\frac{-\frac{{\bf x}_i\pmb{\mu}_{ik}\pmb{\mu}_{il}{\bf x}_i^\top}{\sqrt{\gamma_i^2+1}}}{\gamma_i^2+1}\left(\sqrt{\gamma_i^2+1}-\alpha_i\right)-\left(\frac{{\bf x}_i\pmb{\mu}_{ik}}{\sqrt{\gamma_i^2+1}}-\bm{y}_{ik}{\bf x}_i\right)\left(\frac{{\bf x}_i\pmb{\mu}_{il}}{\sqrt{\gamma_i^2+1}}-\bm{y}_{il}{\bf x}_i\right)^\top}{\left(\sqrt{\gamma_i^2+1}-\alpha_i\right)^2} \\ 
&& - \frac{d-1}{2}\sum_{i=1}^n\frac{\frac{-\frac{{\bf x}_i\pmb{\mu}_{ik}\pmb{\mu}_{il}{\bf x}_i^\top}{\sqrt{\gamma_i^2+1}}}{\gamma_i^2+1}\left(\sqrt{\gamma_i^2+1}-1\right)-\left(\frac{{\bf x}_i\pmb{\mu}_{ik}}{\sqrt{\gamma_i^2+1}}\right)\left(\frac{{\bf x}_i\pmb{\mu}_{il}}{\sqrt{\gamma_i^2+1}}\right)^\top}{\left(\sqrt{\gamma_i^2+1}-1\right)^2} \\
&& + \frac{d-1}{2}\sum_{i=1}^n\frac{-4\mu_{ik}{\bf x}_i\mu_{il}{\bf x}_i^\top}{\gamma_i^4}.
\end{eqnarray*}

\subsection{Asymptotic normality of the maximum likelihood estimators for the regression models in Section \ref{sec:reg}}

Under certain regularity conditions \citep[see, e.g.,][Theorem 5.39]{vaart1998}, general theory implies that the asymptotic normality, specifically (\ref{eq:asy_normal_sc}) and (\ref{eq:asy_normal_pkb}), hold for the maximum likelihood estimators of the regression models given in Section \ref{sec:reg}.

The observed Fisher information matrices, ${\cal I}_{SC}$ and ${\cal I}_{PKB}$, which appear in (\ref{eq:asy_normal_sc}) and (\ref{eq:asy_normal_pkb}), respectively, can be calculated as follows.
It is straightforward to see that ${\cal I}_{SC}$ takes the form
\begin{align}
{\cal I}_{SC} & = - \frac{1}{n} \frac{\partial^2 \ell_{SC}}{\partial \pmb{\beta} \partial \pmb{\beta}^\top} \nonumber \\
& = \frac{d}{n} \sum_{i=1}^n \frac{ \left( \sqrt{\gamma_i^2+1} - \alpha_i \right) \frac{\partial^2}{\partial \pmb{\beta} \partial \pmb{\beta}^\top} \left( \sqrt{\gamma_i^2 +1} - \alpha_i \right)  - \frac{\partial}{\partial \pmb{\beta}} \left( \sqrt{\gamma_i^2 +1} - \alpha_i \right) \frac{\partial}{\partial \pmb{\beta}} \left( \sqrt{\gamma_i^2 +1} - \alpha_i \right)^\top    }{ ( \sqrt{\gamma_i^2+1} - \alpha_i )^2}. \label{eq:fisher_sc_proof}
\end{align}
The components of the derivatives in (\ref{eq:fisher_sc_proof}), required to calculate ${\cal I}_{SC}$, are provided in Appendix \ref{sec:derivatives_SC}.
Using these components, the derivatives in (\ref{eq:fisher_sc_proof}) can be expressed as
$$
\frac{\partial}{\partial \pmb{\beta}} \left( \sqrt{\gamma_i^2 +1} - \alpha_i \right) = \frac{1}{\sqrt{\gamma_i^2+1}}
 \begin{pmatrix}
 (\pmb{\beta}_1^\top \bm{x}_i ) \bm{x}_i \\
 \vdots \\
 (\pmb{\beta}_{d+1}^\top \bm{x}_i ) \bm{x}_i
 \end{pmatrix} 
 - 
 \begin{pmatrix}
 y_{i1} \bm{x}_i \\
 \vdots \\
 y_{i,d+1} \bm{x}_i
 \end{pmatrix} 
 = \left( \frac{\bm{\mu}_i}{\sqrt{\gamma_i^2+1}} - \bm{y}_i \right) \otimes \bm{x}_i ,
$$
\begin{align*}
\frac{\partial^2}{\partial \pmb{\beta} \partial \pmb{\beta}^\top} \left( \sqrt{\gamma_i^2 +1} - \alpha_i \right) & = \frac{\partial}{\partial \pmb{\beta}^\top} \left( \frac{\bm{\mu}_i}{\sqrt{\gamma_i^2+1}} - \bm{y}_i \right) \otimes \bm{x}_i \\
&= \frac{ \sqrt{\gamma_i^2+1} \frac{\partial }{ \partial \pmb{\beta}^\top} \left( \bm{\mu}_i \otimes \bm{x}_i \right)  - \left( \bm{\mu}_i \otimes \bm{x}_i \right) \frac{\partial}{\partial \bm{\beta}^\top} \sqrt{\gamma_i^2+1}   }{\gamma_i^2+1} \\
& = \frac{ \sqrt{\gamma_i^2+1} \, {\bf I}_{(d+1)p} \otimes \bm{x}_i^\top \otimes \bm{x}_i 
 - ( \bm{\mu}_i \otimes \bm{x}_i ) ( \bm{\mu}_i  \otimes \bm{x}_i )^\top }{\gamma_i^2+1} .
\end{align*}
Substituting these results into (\ref{eq:fisher_sc_proof}), we obtain the observed Fisher information in the form of (\ref{eq:fisher_sc}).

The observed Fisher information ${\cal I}_{PKB}$ for the PKB-based regression model can be derived from (\ref{pkbdreg}) as
\begin{align}
\begin{split} \label{eq:fisher_pkb_proof}
{\cal I}_{PKB} = - \frac{1}{n} \frac{\partial^2 \ell_{PKB}}{\partial \pmb{\beta} \partial \pmb{\beta}^\top} = & - \frac{d+1}{2n} \sum_{i=1}^n \frac{\partial^2}{\partial \pmb{\beta} \partial \pmb{\beta}^\top} \log \left( \sqrt{\gamma_i^2+1} - \alpha_i \right) \\
& - \frac{d-1}{2n} \sum_{i=1}^n \frac{\partial^2}{\partial \pmb{\beta} \partial \pmb{\beta}^\top} \left\{ \log \left( \sqrt{\gamma_i^2+1} -1 \right) - \log  \gamma_i^2  \right\}.
\end{split}
\end{align}
Note that the calculations for ${\cal I}_{SC}$ and the results in Appendix \ref{sec:derivatives_PKB} imply
$$
\sum_{i=1}^n \frac{\partial^2}{\partial \pmb{\beta} \partial \pmb{\beta}^\top} \log \left( \sqrt{\gamma_i^2+1} - \alpha_i \right) = \frac{n}{d} \, {\cal I}_{SC},
$$
\begin{align*}
\frac{\partial^2}{\partial \pmb{\beta} \partial \pmb{\beta}^\top} \log \left( \sqrt{\gamma_i^2+1} -1 \right) & = 
\frac{ \left( \sqrt{\gamma_i^2+1} - 1 \right) \frac{\partial^2}{ \partial \pmb{\beta} \partial \pmb{\beta}^\top } \sqrt{\gamma_i^2+1}  - \frac{\partial}{\partial \pmb{\beta}} \sqrt{\gamma_i^2+1} \frac{\partial}{\partial \pmb{\beta}^\top} \sqrt{\gamma_i^2+1} }{ \left( \sqrt{\gamma_i^2+1} -1 \right)^2 }  \\
 & = \frac{ \left( \sqrt{\gamma_i^2+1} -1 \right) {\bf I}_{d+1} \otimes \bm{x}_i^\top \otimes \bm{x}_i - (\gamma_i^2+1)^{-1} ( \bm{\mu}_i \otimes \bm{x}_i ) ( \bm{\mu}_i  \otimes \bm{x}_i )^\top }{ \left( \sqrt{\gamma_i^2+1} -1 \right)^2 },
\end{align*}
\begin{align*}
\frac{\partial^2}{\partial \pmb{\beta} \partial \pmb{\beta}^\top} \log \gamma_i^2 & = 
\gamma_i^{-4} \left(  \gamma_i^2 \frac{\partial^2}{ \partial \pmb{\beta} \partial \pmb{\beta}^\top } \gamma_i^2   - \frac{\partial}{\partial \pmb{\beta}} \gamma_i^2 \frac{\partial}{\partial \pmb{\beta}^\top} \gamma_i^2 \right)   \\
 & = \gamma_i^{-4} \left\{ 2 \gamma_i^2 {\bf I}_{d+1} \otimes \bm{x}_i^\top \otimes \bm{x}_i - 4 ( \bm{\mu}_i \otimes \bm{x}_i ) ( \bm{\mu}_i \otimes \bm{x}_i )^\top \right\}.
\end{align*}
Substitution of these results into (\ref{eq:fisher_pkb_proof}) leads to the observed Fisher information expressed as (\ref{eq:fisher_pkb}).


\clearpage
\bibliographystyle{apalike}
\bibliography{biblio}

\end{document}